\documentclass[superscriptaddress,twocolumn,nofootinbib]{revtex4}
\usepackage{graphicx}
\usepackage{color}
\usepackage{enumerate}
\def\be{\begin{equation}} 
\def\ee{\end{equation}} 
\def\veps{\varepsilon}
\renewcommand{\vec}[1]{\mbox{\boldmath $#1$}}

\begin{document}

\title{A microscopic model for spontaneous 
fission: \\
validity of the adiabatic approximation}

\author{K. Hagino}
\affiliation{ 
Department of Physics, Kyoto University, Kyoto 606-8502,  Japan} 

\author{G.F. Bertsch}
\affiliation{ 
Department of Physics and Institute of Nuclear Theory, Box 351560, 
University of Washington, Seattle, Washington 98915, USA}


\begin{abstract}
We investigate microscopically the tunneling dynamics in 
spontaneous fission of atomic nuclei. To this end, we employ 
a schematic solvable model with a pairing-plus-quadrupole interaction. 
The spontaneous decay of a system 
is simulated by introducing a small imaginary part 
to the energy of a fission doorway state. 
We show that the many-body Hamiltonian can be reduced to an effective 
2$\times$2 Hamiltonian, from which one can 
derive a simple approximate formula for the decay 
width. 
We particularly investigate the applicability of the adiabatic approximation, 
which has often been used in the literature. 
With typical value of the parameters, 
we find that the adiabatic 
approximation 
may underestimate the decay width
by orders of magnitude, depending on the number of orbital transitions.
\end{abstract}


\maketitle

\section{Introduction}

Nuclear fission is a primary decay mode of heavy nuclei. It 
plays an important role in a diversity of phenomena, 
including nuclear technology, syntheses of superheavy elements, and 
r-process nucleosynthesis. While there has been much recent 
progress in the  theory \cite{D19}, 
its microscopic understanding is still far from complete.
An adequate quantum description not only has to deal with the very large
changes in shape, but also with huge number 
of many-body configurations that are involved in the transition.  
One of the ultimate goals of low-energy nuclear 
theory is 
to develop a microscopic framework to describe this complex dynamics.
For that purpose, one would need an efficient 
truncation scheme in order to handle the problem within a 
manageable computation time. 

Given this situation, it may be 
useful to consider solvable microscopic models to
test the reliability of
the approximations in current use, and perhaps even to suggest new
approximation schemes.  A good model should be simple,  
yet should contain the essential features of large-amplitude quantum
dynamics. 
One of us (G.F.B.) has proposed a model along these lines, reported in Ref.
\cite{Bertsch19}.  There the model was
applied to induced fission, that is, fission in a nucleus excited about the
fission barrier.   It was demonstrated in that paper that the  
branching ratio in the competition between the fission and the 
capture reaction is sensitive to the character of the residual 
interaction. 

In this paper, we apply the model
to spontaneous fission, but with Hamiltonian parameters adapted to
that process. Here barrier penetration 
plays a decisive role and the barrier height can be controlled by
one of the parameters. 
We shall apply the model to investigate the accuracy of the adiabatic 
approximation, which has often been employed in microscopic 
calculations for nuclear fission. 

The paper is organized as follows. In Sec. II, we introduce the 
model Hamiltonian used in our investigations. 
In Sec. III, by numerically diagonalizing the Hamiltonian matrix, we 
investigate the dependence of the decay width on 
several parameters in the model. In Sec. IV, we introduce the 
adiabatic approximation and discuss its applicability. 
We then summarize the paper in Sec. V. 
In Appendix A, we discuss 
an alternative way to 
solve the model Hamiltonian using 
a time-dependent approach. 

\section{Model Hamiltonian}

The model Hamiltonian introduced in Ref. \cite{Bertsch19} reads
\begin{equation}
\hat H = \sum_{k=0}^{N_{\rm orb}-1}  \varepsilon_k \hat n_k + 
v_Q \hat Q \hat Q +\sum_{k,k'} v_{kk'} \hat P^\dagger_k \hat P_{k'}
\label{hamiltonian}
\end{equation}
where  
$\hat{n}_k=a_k^\dagger a_k+a_{\bar{k}}^\dagger a_{\bar{k}}$ is the 
number operator for orbital $k$ including its time-reversed partner
$\bar{k}$,   and $\epsilon_k$ is the single-particle energy of the  orbital
and its partner. 
The operator $\hat{Q}$ represents a shape-dependent fields such as 
the quadrupole operator.  It is  defined as $\hat{Q}=\sum_kq_k\hat{n}_k$.
Finally, the operator
$\hat{P}_k^\dagger=a^\dagger_ka^\dagger_{\bar{k}}$ creates a pair in
one of the orbitals. 

As in Ref. \cite{Bertsch19}, we study the model in a configuration space
containing $N_p=6$  particles
in $N_{\rm orb} = 6$ orbitals.  
The orbitals are grouped by shape;  the first three favor the ground-state
shape and the last three the shape associated with the scission
configuration.  This is implemented with orbital quadrupole moments 
$q_k=-1$ for the first three and 
$q_k=1$ for for the last three.
Labeling the orbitals as ($k=0,1,\cdots, 5$), we
set the single-particle energies $\epsilon_k$ 
as 
\begin{equation}
\varepsilon_k=(k \bmod (N_{\rm orb}/2))\,  \varepsilon_0 
\label{sp_energy}
\end{equation}
where $\varepsilon_0$ is the single-particle level spacing.  
Table I summarizes the single-particle energies
and the quadrupole moment for each orbital.
In  Ref.
\cite{Bertsch19} there was added a small random energy to break some
unwanted degeneracies.  In the application to barrier penetration,
the only
degeneracy of consequence is between the end configurations.  We
shall deal with this by introducing a shift $\Delta$ to the diagonal energy
of the prescission configuration.  
\begin{table}
\caption{The single-particle energy $\epsilon_k$ and the 
quadrupole moment $q_k$ for single-particle orbitals in the
$(N_{\rm orb},N_p) = (6,6)$ model.
The single-particle energies are given in units of $\epsilon_0$ 
}
\begin{tabular}{|c|cccccc|}
\hline
\hline
$k$ & 0 & 1 & 2 & 3 & 4 & 5  \\
\hline
$\epsilon_k$ & 0 & 1 & 2 & 0 & 1 & 2 \\
$q_k$ & $-1$ & $-1$ & $-1$ & \,\,1 & \,\,1 & \,\,1 \\
\hline
\hline
\end{tabular}
\end{table}
 
The two-body interaction is taken 
to purely pairing in the form 
\begin{equation}
v_{kk'}=-G(1-\delta_{k,k'}). 
\label{pairing}
\end{equation}
With this interaction the seniority of the wave function is a good quantum
number.  Note that the diagonal matrix elements of the pairing
Hamiltonian are set to zero.  This does not affect the wave function
and permits a better separation between the diagonal and
off-diagonal parameterization in the Hamiltonian.
We
consider below only the seniority-zero subspace, namely configurations
having 3 pairs in the 6 orbitals.  The total number of configurations in 
the space is 
\be
N_{\rm conf} = {N_{\rm orb} \choose N_p/2} = 20. 
\ee

\begin{figure} [tb]
\includegraphics[width=1.0\columnwidth,clip]{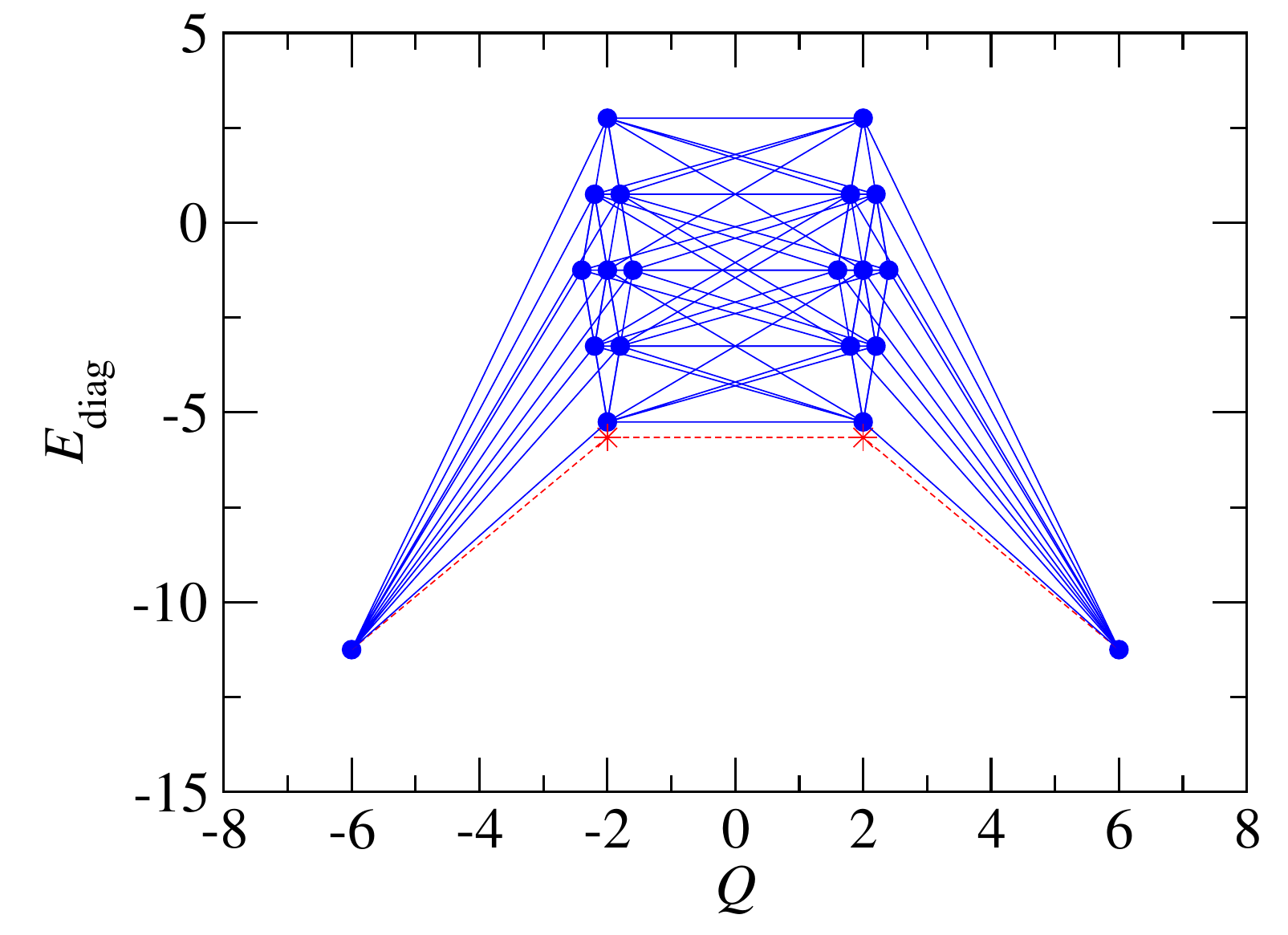}
\caption{
\label{fig1}
Diagonal energies of the configurations in the $(N_{\rm orb},N_{p})=(6,6)$ model
space with shape parameter Eq. (\ref{vq})  are shown as the filled  
circles.
The off-diagonal interaction connects configurations joined by lines. 
Stars show energies of adiabatic intermediate states; see Sec. \ref{testing} for
details.
}
\end{figure}

We now discuss the choice of parameters.
Physically the
most relevant parameters in spontaneous fission by pairing
dynamics is the barrier height $V_B$ and the pairing gap
$\Delta_{\rm BCS} $. Typical physical values in the actinide nuclei 
are $V_B\approx 5$ MeV, $\veps_0 \approx 0.5 $
MeV, and
$\Delta_{\rm BCS} \approx 0.8  $ MeV.   The model has only two dimensionless
parameters, $v_Q/\veps_0$ and  $G/\veps_0$, besides the size parameters
$N_{\rm orb}$ and $N_p$.  We will eventually vary the parameters, but for now
take the shape dependence as
\be
\label{vq}
 v_q = -0.3125 \,\veps_0.
\ee
The resulting spectrum of energies of the individual configurations is 
shown in Fig. \ref{fig1}.  

There are four sets of configurations distinguished by 
their expectation values of the $Q$ operator, with one
configuration in the extreme sets and nine in the
interior sets.
The leftmost configuration is the main component of the
ground state, while the rightmost one represents a doorway
to the fission channels.  These two configurations are
degenerate in the model as so far presented.
The energy gap
between the ground state configuration and the lowest states in
the interior configurations is $V_B = 6\veps_0$. This is lower than a typical
physical barrier, but as mentioned earlier we will consider parameter
variations over a broad range.  

For the pairing
interaction strength $G$, we compare with a physical pairing strength 
via the BCS approximation to the pairing gap.  Here we carry out
the BCS calculation in a space of $N_{\rm orb}$ orbitals occupied by
$N_p/2$ pairs.  For this calculation we assume that the orbital energies
are evenly spaced by an energy difference $\veps_0$.  
For $N_{\rm orb}=6$ and $N_p=6$, 
an
interaction strength of  $G / \veps_0=0.691$ produces 
the BCS gap of
$\Delta_{\rm BCS} / \veps_0=3/2 $. 
This is close to the above rough estimate and we carry out
the Hamiltonian calculations with it.
Finally, we modify the diagonal energy $E_d$ of the doorway configuration by
adding a small imaginary part $-i\Gamma_d/2$ and a small real part $\Delta$,
$E_d = E_g + \Delta -i\Gamma_d/2$ where $E_g$ is the energy of the
leftmost configuration.  The resulting non-Hermitian Hamiltonian is
diagonalized to obtain a spectrum of decaying states.  The decay width
is given by 
\be
\Gamma_f = -2 {\rm Im}\,E_{\rm gs},  
\ee
where $E_{gs}$ is the eigenenergy of the state having the largest
component of the leftmost configuration.
The procedure will fail if the two end states are degenerate, because
then there will be two candidates having 
nearly equal amplitudes for
the configuration.  We therefore have to understand the dependence
of the calculated width on their (real) splitting $\Delta$. As we
show below, the lack of specific knowledge of $\Delta$ is not
an obstacle to assess the adiabatic approximation.

\section{Decay Width}

We now examine the dependence of the ground state decay width 
on the doorway width.
\begin{figure} [tb]
\includegraphics[width=1.0\columnwidth,clip]{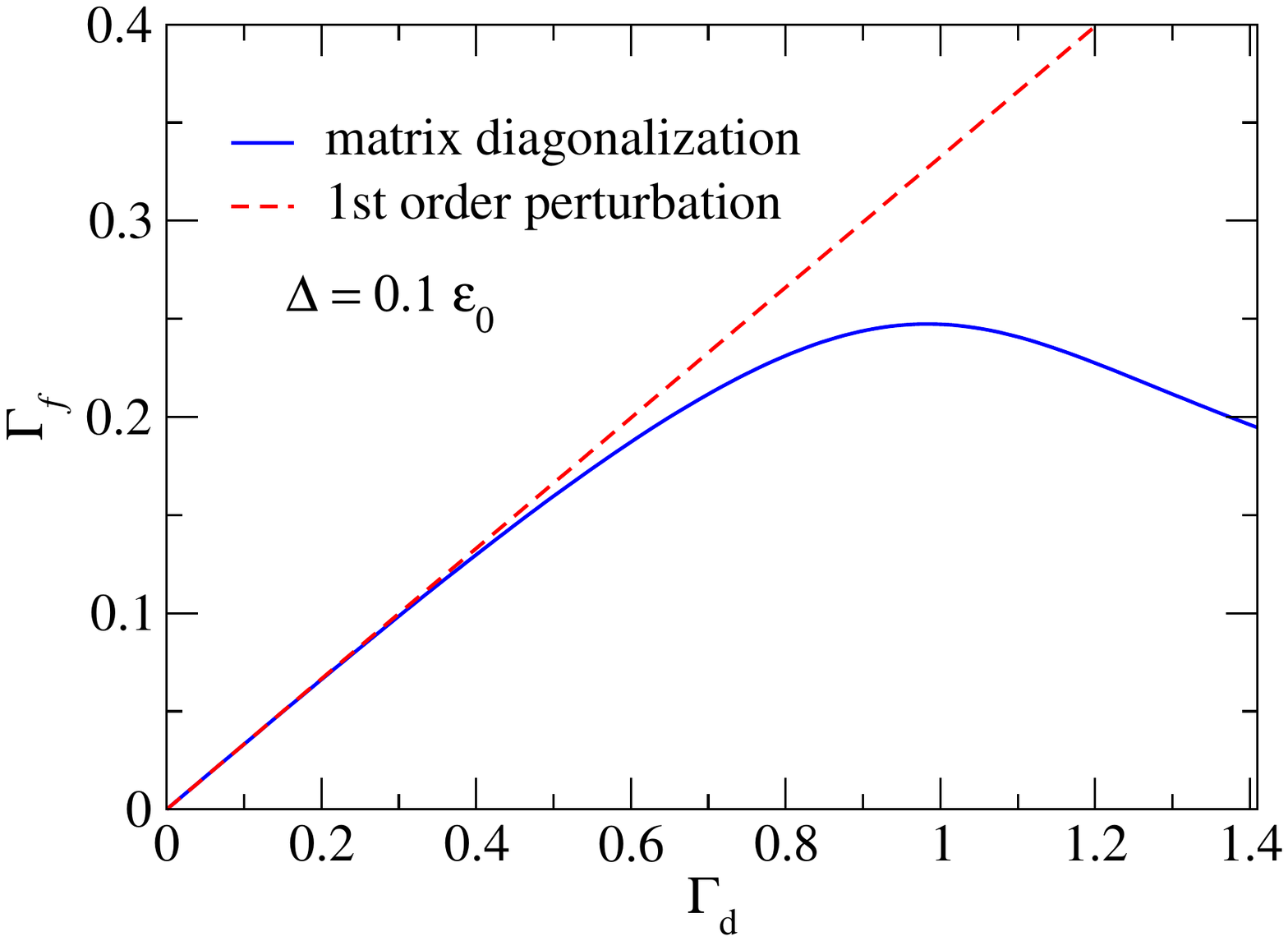}
\caption{
\label{fig2}
The decay width of spontaneous fission, $\Gamma_f$, 
as a function of the 
width of the fission doorway state, $\Gamma_d$. These widths are 
given in units of $\epsilon_0$. 
The solid line is obtained by diagonalizing the non-Hermitian 
Hamiltonian for a system with 3 pairs in 6 orbits, while the dashed 
line is the result of the first order perturbation theory with 
respect to $\Gamma_d$. }
\end{figure}
This is shown in Fig. \ref{fig2} for the offset $\Delta = 0.1$
in the doorway energy.  One sees that $\Gamma_f$ first rises
linearly with $\Gamma_d$, in accord with the first-order
perturbation theory formula  
\be
\label{fop}
\Gamma_f =   \Gamma_d |\phi_d|^2  \,\,\,{\rm (perturbative)}
\ee
where $|\phi_d|^2$ is the probability of the doorway configuration in the
unperturbed ground state.  The region of validity of Eq.
(\ref{fop}) is not broad enough
for our purposes and we do not consider it further. 
Note that $\Gamma_f$ saturates at
larger $\Gamma_d$ and then decreases.
The decrease may be analogous
to the phenomenon of super-radiance discussed e.g., in Ref.
\cite{Auerbach11}. 

Since the offset is important to the calculation, it has to be
fixed when comparing different Hamiltonian approximations. 
This can be achieved by reducing the Hamiltonian matrix to a
2$\times$2 matrix containing only the two end configurations. 
To achieve this, we divide
the configuration space into three parts: the
unperturbed ground state, the fission doorway, and all the
interior configurations as the third part.  Let us call the
Hamiltonian for the interior configurations $H_b$ (for
``barrier'').  The matrix elements coupling $H_b$ to the end
configurations will be designated $\vec{v}_g$ and $\vec{v}_d$,
where the bold-face type is a reminder that these are vectors with
the same dimension as $H_b$.  The Hamiltonian to be diagonalized
has the form
\begin{equation}
H=\left(
\matrix{
E_{g} & \vec{v}_g^T & 0 \cr
\vec{v}_g  & H_b & \vec{v}_d \cr
0 & \vec{v}_d^T & E_g+ \Delta -i\Gamma_d/2 \cr}
\right).
\end{equation}
Here
$E_g$ and $E_g+\Delta $ are the energies of the unperturbed ground state 
and the doorway state, respectively. 
The eigenvector for the decaying state satisfies 
the equation
\begin{equation}
\left(
\matrix{
E_{g} & \vec{v}_g^T & 0 \cr
\vec{v}_g  & H_b & \vec{v}_d \cr
0 & \vec{v}_d^T & E_{g}+ \Delta -i\Gamma_d/2 \cr}
\right)
\left(
\matrix{
\phi_g \cr
\vec{\phi}_b \cr
\phi_d \cr}
\right)
=
E_{gs} 
\left(
\matrix{
\phi_g \cr
\vec{\phi}_b \cr
\phi_d \cr}
\right).
\end{equation}
If one knew the complex ground state energy $E_{gs}$, this equation could be 
solved for $\vec{\phi}_b$ as 
\begin{equation}
\vec{\phi}_b=(E_{gs}-H_b)^{-1}(\vec{v}_g\phi_g+\vec{v}_d\phi_d).
\end{equation}
Substituting this to the original eigenvalue equation, one finds 
\begin{equation}
H_{\rm eff}
\left(
\matrix{
\phi_g \cr
\phi_d \cr}
\right)
=
E_{gs}
\left(
\matrix{
\phi_g \cr
\phi_d \cr}
\right), 
\end{equation}
with 
\begin{eqnarray}
&&H_{\rm eff} 
=\left(
\matrix{
E_g+v_{{\rm eff},gg}  & v_{{\rm eff},gd}
 \cr 
v_{{\rm eff},dg}
& E_g + \Delta-i\frac{\Gamma_d}{2}+v_{{\rm eff},dd}
\cr }
\right), 
\label{effH0}
\end{eqnarray}
and 
\be
\label{exact}
v_{{\rm eff},ij} = \vec{v}_i^T(E_{gs}- H_b)^{-1}\,\vec{v}_j. 
\ee
The reduction of the problem to the 2$\times$2 effective
Hamiltonian is exact as long as the eigenenergy $E_{gs}$ is
correct.  One can derive a simpler approximate Hamiltonian
assuming that the fission barrier is much higher than other energies
in the model.  If 
$E_{gs}$ is close to the unperturbed ground state energy, $E_g$,
and the imaginary part is also small, we may
assume $E_{gs}\approx E_{g}$ in evaluating
$(E_{gs}-H_b)^{-1}$. The second-order terms in the diagonal entries
to the Eq. (\ref{effH0}) are also small.  In fact they are equal
for the Hamiltonian Eq. (\ref{hamiltonian}).  In effect, the diagonal
terms only produce a
shift in the total energy which can be ignored if it is small compared
to $V_B$. 
The resulting approximate 
effective Hamiltonian can be written
 \begin{equation} H'_{\rm eff}  =\left( \matrix{  E_g  &  
v_{\rm eff} \cr v_{\rm eff} &  E_g +\Delta 
-i\frac{\Gamma_d}{2} \cr } \right),  \label{effH} 
\end{equation} with  %
\begin{equation}
v_{\rm eff} 
=v_{{\rm eff},gd}
\label{v_eff}
\end{equation}

To assess the accuracy of the approximations we made to derive
Eq. (\ref{effH}), we
compare with the exact second-order terms, Eq. (\ref{exact}).
For parameters $G$ and $v_Q$ given in Sec. II 
in the numerical Hamiltonian, the second-order contribution to diagonal
energies is  $-0.803\veps_0$, which is indeed small compared to $V_B$. 
The effective 
interaction between the two end configurations is found to be
$v_{\rm eff} = -0.260\,\veps$.  The corresponding quantity under
the approximation $E_{gs} \rightarrow E_g$ is 
$v_{\rm eff}= -0.348\,\veps$.  Although the
difference from the exact is larger than we would like, it will
factor out of the quantities we will use to test the adiabatic treatment. 

\begin{figure} [tb]
\includegraphics[width=1.0\columnwidth,clip]{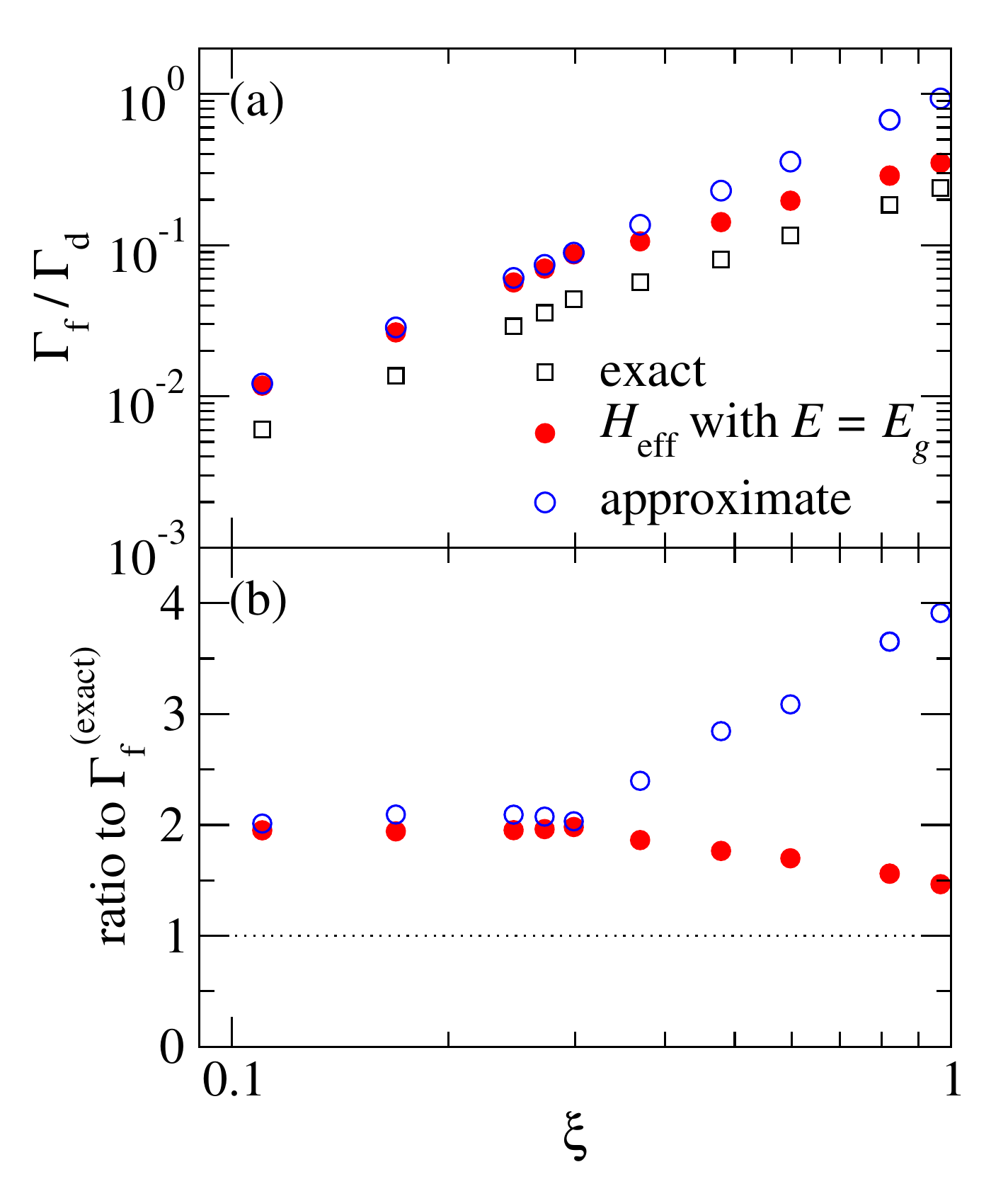}
\caption{
\label{fig3}
The decay width obtained with several methods. It is plotted as
a function of the parameter 
$\xi\equiv -v_{\rm eff}/\sqrt{\Delta^2+\Gamma_d^2/4}$, 
where $v_{\rm eff}$ is the
off-diagonal coupling strength in the effective Hamiltonian, Eq.
(\ref{effH}). For the exact results, shown as the open squares,
the abscissa is the real part of $\xi$. The filled circles are
obtained by the approximation $E_{gs}\rightarrow E_b$ in the
effective Hamiltonian Eq. (\ref{effH0}).  
The open circles show the results of
the approximate formula, Eq. (\ref{width_pert}).  All energies
are measured in units of $\epsilon_0$.
}
\end{figure}

We shall consider one additional approximation for calculating ratios.
Under physical conditions, $v_{\rm eff}$ will be small 
compared to the other energies in Eq. (\ref{effH}),
permitting one to estimate $E_{gs}$ by
second-order perturbation theory.
This leads to 
\begin{equation}
E_{gs} \approx  E_g+ \frac{v_{\rm eff}^2}{-\Delta +i\Gamma_d/2}, 
\end{equation}
from which one obtains the decay width 
\begin{equation}
\Gamma_f \approx \frac{\Gamma_d \,v_{\rm eff}^2}{\Delta^2+\Gamma_d^2/4}.
\label{width_pert} 
\end{equation}
This subsumes the dependence on $v_{\rm eff},\Delta$ and $\Gamma_d$
in a single dimensionless parameter
\be  
\xi\equiv -v_{\rm eff}/\sqrt{\Delta^2+\Gamma_d^2/4}.
\ee 
We can now assess the accuracy of the approximations as a 
function of $\xi$.  The comparison with the exact is
shown in Fig. \ref{fig3} as a function of $\xi$, but
computed for a 
range $[0.2-3.0]\veps_0$ of the $\Delta$ and $\Gamma_d$
parameters. 

One can see that the approximation to replace $E_{gs}$ with $E_g$ in the 
effective Hamiltonian 
is independent of $\xi$ 
over the entire parameter
range.  The perturbative formula, 
Eq. (\ref{width_pert}), is fairly accurate for $\xi < 0.3$.
For spontaneous fission, we can assume that $v_{\rm eff}$ is much smaller than
$\Delta$ and $\Gamma_d$.   Thus, we can discuss the 	
dynamics of spontaneous fission
with the formula
\be
\Gamma_f \approx \Gamma_d \xi^2.
\ee
We shall apply this to the adiabatic 
approximation in the next section. 

\section{Testing the adiabatic approximation}
\label{testing}
Most of the theoretical calculations for spontaneous and  
low-energy fission 
found in literature have 
been based on the adiabatic approximation, see e.g., 
Refs. \cite{Goutte2005,Warda2012,Jhilam2013,Staszczak2013,
Shunck2016,Nakatsukasa2016}\footnote{Nonadiabatic
effects have also been discussed in the literature, see e.g.
Refs. \cite{SW75,BNV83,BS83,Bernard2011,Bulgac2016}.}. 
For instance, when a potential surface for fission is constructed 
microscopically, one often employs the constrained Hartree-Fock method 
to minimize the energy for a given nuclear shape. Thus only the
local ground state at each deformation is taken into consideration. 

A nice feature of the schematic model presented here is that it
can be used to assess the validity of such approximations.  In
the present model, the adiabatic approximation is implemented by
first diagonalizing the model Hamiltonian within subspaces
of fixed  $Q$.  Then one constructs 
a new basis taking only the  lowest energy state of each $Q$
value. In the case of 3 pairs in 6 orbits, this approximation
reduces the dimension of the Hamiltonian from 20 to 4; the
states in the reduced basis have
$Q_= -6, -2, 2$, and 6.  The eigenenergies of the four
states are shown by stars in Fig.  1.  We note that the reduced
Hamiltonian is nothing but the discrete model used in the pair
hopping model\footnote{The model has also been applied recently
to $\alpha$ decays
\cite{Rissanen14,Clark18,Clark19}.}  \cite{Barranco90,BF91}.   
In the pair hopping
model, the discrete-basis representation of the Hamiltonian is
transformed to a Schr\"odinger-like equation, from which the
inertia parameter for fission is deduced.  In this paper, we
instead diagonalize the reduced Hamiltonian matrix as it is and
compute the decay width from the eigenenergy of the ground
state.
\begin{figure} [tb]
\includegraphics[width=1.0\columnwidth,scale=0.5,clip]{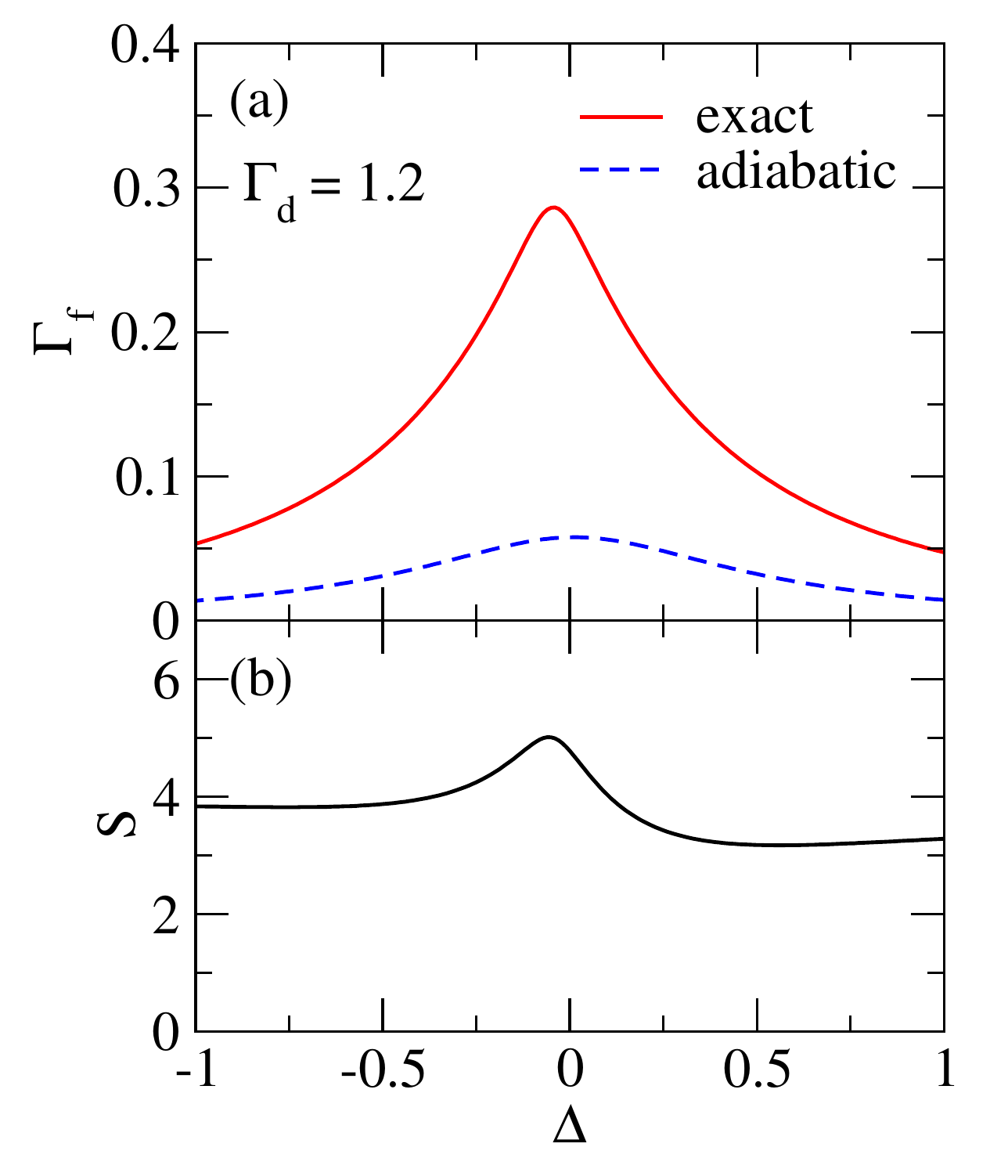}
\caption{
\label{fig4} The Upper panel compares the fission width 
from the exact diagonalization of the Hamiltonian matrix (solid line) 
with the adiabatic approximation (the dashed line) as a function
of the offset $\Delta$ and with $\Gamma_d=1.2\epsilon_0$. 
Energies are measured in units of $\epsilon_0$. 
The lower panel shows the ratio $S=
\Gamma_f({\rm exact})/\Gamma_f({\rm adiabatic})$.
}
\end{figure}

The upper panel of Fig. \ref{fig4} shows a comparison of 
the decay width from the exact diagonalization 
of the original Hamiltonian matrix (the solid line) 
to the width in the adiabatic approximation (the dashed line) as
a function of the offset $\Delta$, taking $\Gamma_d=1.2 \epsilon_0$.
The ratio between the two is plotted in the lower panel. 
One can see that the adiabatic approximation suppresses the 
decay width by almost a factor of 4.  One can extract
a suppression factor $S$ without carrying out the full width calculation,
using instead the perturbative estimate
\be
\label{ratio}
S = \frac{\Gamma_f}{\Gamma_f'} = \left(\frac{v_{eff}}{v_{eff}'}\right)^2
\ee 
where the primed quantities are the adiabatic values.
This yields $S = 4.79$ with 
$v_{\rm eff}=-0.348 \epsilon_0$ and $v'_{\rm eff}=-0.159 \epsilon_0$.

When the barrier Hamiltonian, $H_b$, as well as the vectors $\vec{v}_g$ and 
$\vec{v}_d$ in Eq. (\ref{v_eff}) are 
expressed with the adiabatic basis, that is, the eigenstates of 
the Hamiltonian for a fixed value of quadrupole moment, $Q$, 
the lowest energy configurations indeed lead to 
the dominant contribution in Eq. (\ref{v_eff}). 
However, contributions from the other configurations are 
not negligible, and they may provide an important contribution as a whole 
when the number of configurations is not small. 
In order to demonstrate this, Fig. 5 shows the real 
part of the overlap amplitude 
between the 
ground state of the original 20$\times$20 Hamiltonian with 
$\Gamma_d=1.2 \epsilon_0$ and each of the 
adiabatic state as a function of the energy of the adiabatic states. 
The upper and the lower panels are for $Q=-2$ and 2, respectively. 
Here, the overlap amplitude $O_k(Q)$ is defined as 
\begin{equation}
O_k(Q)=\frac{\langle \phi_k(Q)|\psi_{\rm gs}\rangle}
{\sqrt{\langle \tilde{\psi}_{\rm gs}|\psi_{\rm gs}\rangle}},
\label{overlap}
\end{equation}
where $|\phi_k(Q)$ is the $k$-th eigenstate of the sub-Hamiltonian 
spanned by the configurations with $Q$. $|\psi_{\rm gs}\rangle$ and 
$|\tilde{\psi}_{\rm gs}\rangle$ are the ground state of the 
non-Hermitian Hamiltonian $H$ 
and its Hermitian conjugate, $H^\dagger$, respectively, which are 
normalized as $\langle\tilde{\psi}_{\rm gs}|\psi_{\rm gs}\rangle=1$ \cite{MF52}. 
One can see that the overlap is indeed the largest for the lowest 
energy configuration for each $Q$. The overlap amplitude 
with the next two states 
are smaller than the overlap with the lowest energy state 
by a factor of around 2.27 for $Q=-2$ and 2.99 for $Q=2$. 
These are small but not negligible and contribute significantly when 
all the contributions are summed up. 

\begin{figure} [tb]
\includegraphics[width=1.0\columnwidth,clip]{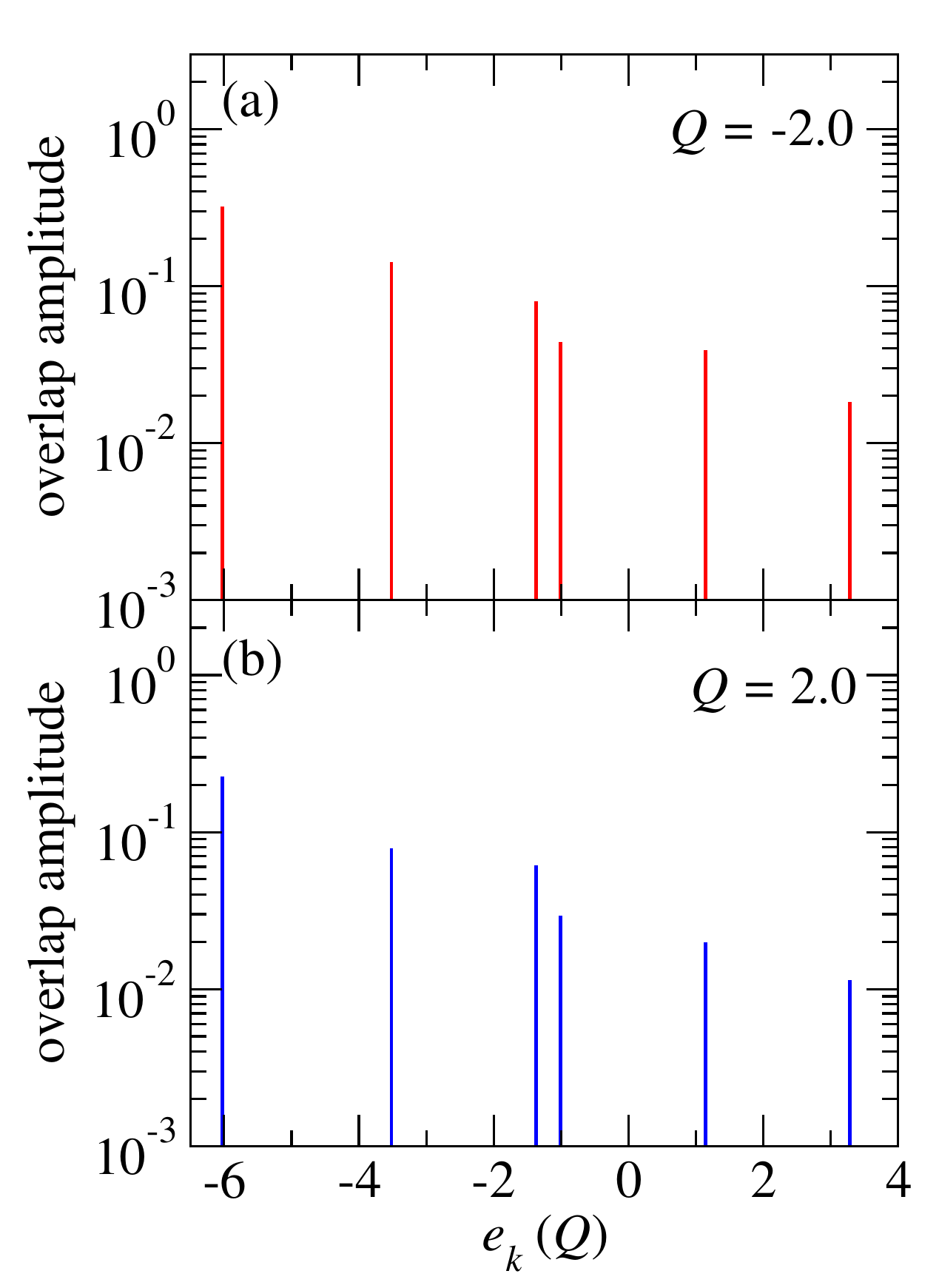}
\caption{
\label{fig5}The real part of the overlap amplitude 
between the ground state of the original 20$\times$20 
Hamiltonian and the adiabatic basis states as a function of the 
energy of each of the adiabatic state. 
The offset and 
the width of the fission doorway state are set to be 
$\Delta = 0.1\epsilon_0$ and 
$\Gamma_d=1.2 \epsilon_0$, respectively. 
The upper panel is for $Q=-2$ while the lower panel is for $Q=2$. 
}
\end{figure}

\begin{table*}
\caption{
Comparision of the square of the effective coupling strength, 
$v_{\rm eff}^2$, with that in the adiabatic approximation, 
$(v^{{\rm ad}}_{\rm eff})^2$, for several parameter sets of the model 
Hamiltonian. 
\label{sensitivity}
}
\begin{tabular}{|c|c|c|c|c|c|c|c|c|}
\hline
\hline
\multicolumn{6}{|c|}{Model} & \multicolumn{3}{c|}{2$\times$2 matrix reduction} \\
\hline
$(N_{\rm orb},N_p)$ & 
$N_{\rm conf}$ & $V_B$ & $v_Q$ & $G$ & $\Delta_{\rm BCS}$ & 
$v_{\rm eff}^2$ & $(v^{{\rm ad}}_{\rm eff})^2$ & ratio \\
\hline
(6,6) & 
20 & 6 &$-5/16$ & 0.691 & 1.5 
& 0.121 & 0.0253 & 0.209 \\
(6,6) & 20 & 6 & $-5/16$ & 1.44 & 4.0  
& 21.24 & 21.30 &
1.003 \\
(6,6) & 20 & 10 &$-7/16$ & 0.691 
& 1.5 
& 1.23$\times 10^{-2}$ & 2.17$\times 10^{-3}$ & 0.177\\
(8,8) & 70 & 6 & $-7/32$ & 0.585 & 1.5 & 1.18 & 2.75$\times 10^{-2}$ & 2.32$\times 10^{-2}$\\
(10,10) & 252 & 6 & $-3/16$ & 0.521 & 1.5 
& 7.85 & 1.75$\times 10^{-2}$ & 2.22$\times 10^{-3}$\\
\hline
\hline
\end{tabular}
\end{table*}

It is interesting to recall that the adiabatic approximation 
gives the upper limit of the tunneling probability in the problem of 
an external potential barrier with a fixed incident energy \cite{BT98,HT12}. 
This is not the case with our Hamiltonian; here the adiabatic
approximation gives a much smaller decay rate. In fact the
models are so different that it should not be surprising that
even qualitative features are affected.

Next we examine the sensitivity of the adiabatic approximation
to physical parameters in the model: the pairing condensate
$\Delta_{\rm BCS}$, the barrier height $V_B$, and the size of the
configuration space $N_{\rm conf}$.  Table \ref{sensitivity}
compares $v_{\rm eff}^2$ calculated with the full Hamiltonian matrix   
and with the adiabatic approximation.  Their ratio is
approximately equal to the ratio of decay widths according to Eq.
(\ref{ratio}).  One sees that the adiabatic approximation
becomes much better when the pairing strength is increased. 
Indeed, in the limit  $\veps_0/G \rightarrow 0 $ the
adiabatic treatment is exact.

However, the adiabatic approximation is seen to fail badly as a function of 
$N_{\rm conf}$, taking the physical parameters at the nominal values.  
The three cases in the table are $(N_{\rm orb},N_p) = (6,6), (8,8)$, and
(10,10), and the error in the adiabatic approximation
is about of factor of 0.2,0.02, and 0.002, respectively.  In these
spaces the number of interior sets of given $Q$ are 2,3,and 4.  To traverse
the space from one end to the other requires an additional jump between
$Q$ values for addition set.  Thus, it appears that the error cost for
each $Q$ transition is about one order of magnitude.  This would be huge
for a space large enough to represent the number of transitions required 
for an actinide nucleus.

\section{Summary}

We have presented a schematic model for spontaneous fission
very different in spirit to previous theory.  Our model is
anchored in the Configuration Interaction (CI) framework
of many-particle quantum mechanics.  In contrast,the
previous theory followed the picture of a particle 
tunneling under a one- or few-dimensional barrier.
Since the CI space in the model is tiny compared to huge
space needed for a quantitative theory, our findings are
at best qualitative.

One interesting finding supporting previous studies \cite{ro17} is
the strong dependence of spontaneous decay rates on the
strength of the pairing interaction.  

The main focus of our study is the validity of the adiabatic
approximation for the intermediate states, and we found a serious
problem in that method.  The approximation was 
a factor of 5  
too small for Hamiltonian containing two stages
of intermediate states, and about one order of magnitude smaller
for each additional stage.
In the physical problem, there are about 20
intermediate states \cite{be19c}, so clearly one needs to
rethink how to deal with approximations that reduce the
dimensionality to one or a few variables.   

There are several directions that can be explored to make
the model more realistic.  One improvement over the
adiabatic approximation is to simulate the least-action 
treatment which has been shown to increase the decay rates
by several orders of magnitude \cite{ro18}.  Instead of the
ground state wave function in the subspaces, the model would
artificially increase the pairing strength in calculating the
intermediate-state wave functions.  The question then arises of how much to increase the
pairing strength.  As a naive possibility, we have examined the change in decay
rate for the $(6,6)$ model space, taking the pairing strength for the two intermediate wave functions that
maximizes the final decay rate.  Indeed, the result comes out close to
the exact, but so far we have not found a good justification for the
procedure.

Another problem of the model is its oversimplification of the
exit from the tunneling region.
In 
$\alpha$-particle decay, the exit from the tunneling region is
straightforward because the barrier is due to
the Coulomb potential field, and the dynamics is essentially
one-dimensional barrier penetration.
In contrast, spontaneous fission produces many
different final states suggesting that there are many doorways
to fission.  Thus, the  model is unrealistic in having only a single doorway.
But the same criticism can be made for the traditional treatment
of spontaneous fission.  

Finally, the model is oversimplified in that the single-particle space
only contains nucleons of the same isospin, that is, all neutrons 
or all protons.  In principle it is straightforward to generalize 
the model to include both species, and the dimension $N_{\rm conf}$ 
remains managable for the $(6,6)$ space in each species.  It is
not clear how this generalization would affect the conclusions.

The ultimate goal is to build more realistic Hamiltonians in the CI
basis.  The configurations can be constructed by the constrained
Hartree-Fock or Hartree-Fock-Bogoliubov method \cite{BF91,BY19}. 
The configurations obtained that way
are not necessarily orthogonal, but that has not posed a significant
problem in other physics fields \cite{nar95}.

\section*{Acknowledgments}
We thank J. Dobaczewski, W. Nazarewicz and other participants in the 
workshop ``Future of Fission Theory'', York, UK (2019) for discussions 
motivating this study. The work of KH was supported in part by 
JSPS KAKENHI Grant Number JP19K03861.

\appendix

\section{Time-dependent approach}

Another way to calculate fission decay rates is through a
time-dependent approach
\cite{SCS94,Talou2000,Maruyama12,Oishi14,Scamps15}. 
This  Appendix applies that method to our model Hamiltonian.
The starting point is defining the initial wave
function at time $t=0$, 
$\psi(0)$. Formally the time-dependent wave function is calculated
with the time evolution operator as
\begin{equation}
|\psi(t)\rangle=e^{-iHt/\hbar}|\psi(0)\rangle. 
\end{equation}
where $H$ is the Hamiltonian.
Using the eigenfunctions of $H$, $|\phi_k\rangle$, and those of $H^\dagger$, 
$|\tilde{\phi}_k\rangle$, the wave function can be expressed as 
\begin{equation}
|\psi(t)\rangle
=\sum_ke^{-iE_kt/\hbar}|\phi_k\rangle\langle\tilde{\phi}_k|\psi(0)\rangle,
\label{time-dep}
\end{equation}
where the $E_k$ are left eigenvalues of the Hamiltonian. 
The survival probability is computed as 
$P_{\rm surv}(t)=|\langle \psi(0)|\psi(t)\rangle|^2$. 
From the survival probability, the decay width may be computed either 
as 
\begin{equation}
\Gamma(t)=-\frac{\hbar}{P_{\rm surv}(t)}\,\frac{dP_{\rm surv}(t)}{dt}
\equiv \Gamma_1(t),
\label{gamma1}
\end{equation}
or as
\begin{equation}
\Gamma(t)=-\frac{\hbar}{t}\,\ln P_{\rm surv}(t)
\equiv \Gamma_2(t). 
\label{gamma2}
\end{equation}
If the survival probability has an exponential dependence,
$P_{\rm surv}(t)\sim e^{-\Gamma t/\hbar}$, 
both formulas yield the 
same decay width, $\Gamma$. 

\begin{figure} [tb]
\includegraphics[width=1.0\columnwidth,clip]{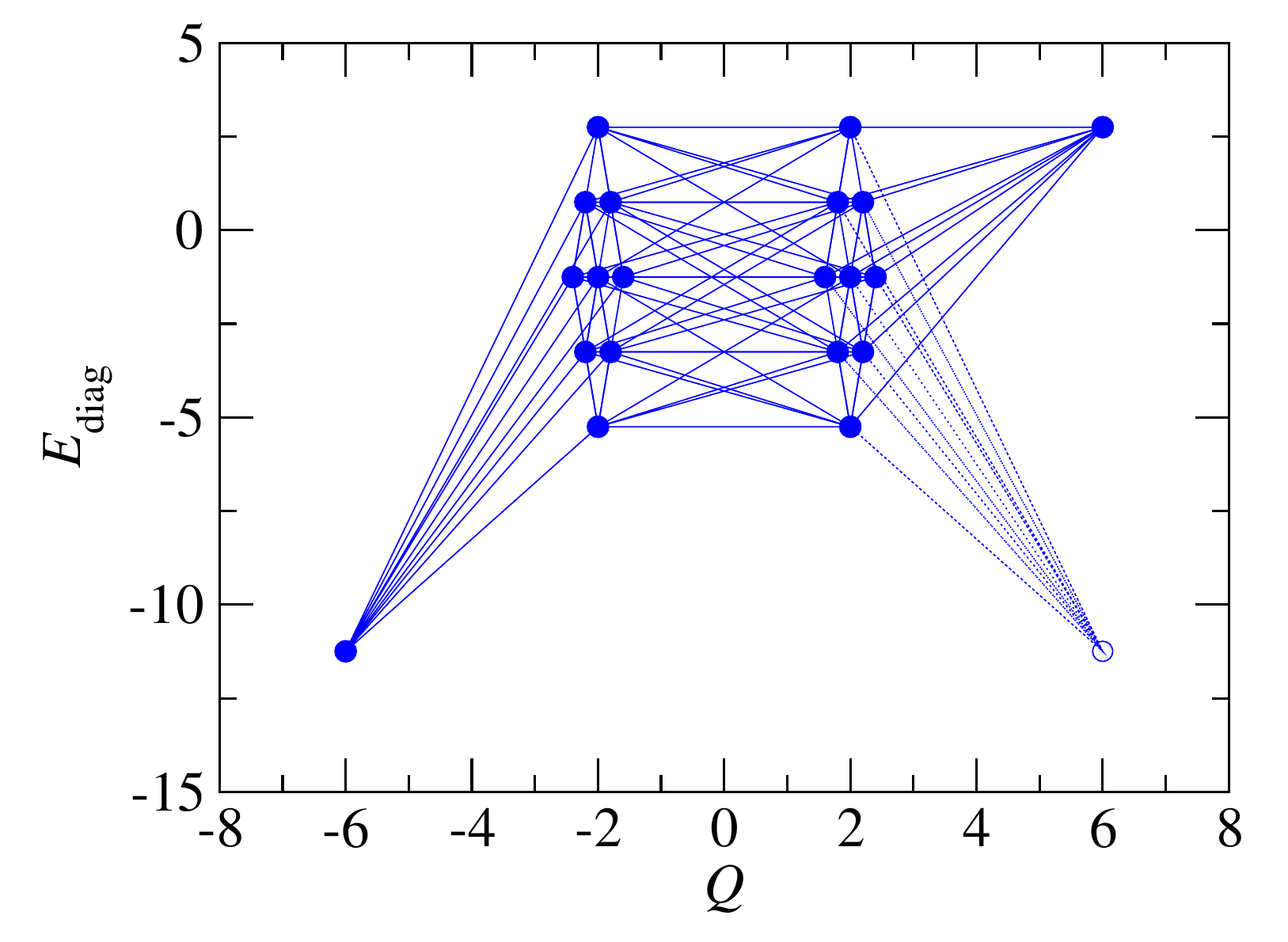}
\caption{
\label{fig6}
A modification of the diagonal energies of the Hamiltonian 
for the system with 3 pairs in 6 orbitals in 
order to construct an initial wave function for the time-dependent 
approach. The diagonal energy outside the fission barrier at $Q=6$ 
is modified from the original value (denoted by the open circle) 
to a constant value which is set to be the same as the 
maximum of the diagonal energy. 
The solid and the dashed lines join configurations that are connected 
by the pairing interaction, for which the dashed lines are used to 
connect to the original energy point. 
}
\end{figure}

In the examples shown below, we take the same non-Hermitian 
Hamiltonian as the one given in Sec. II, taking $\Gamma_d = 1.2 \veps_0$.
When the initial wave function, $\psi(0)$, is taken as
the ground state wave function of the Hamiltonian,
one finds the same fission 
width as that obtained by the diagonalization method
method presented in Sec. II. 

In more realistic applications it is not feasible to compute the
ground state eigenfunction and some approximation to $\psi(0)$ is
introduced. There are three choices that we examine here.
They are:
\begin{enumerate}[i)]
\item the unperturbed ground state configuration, $\phi_g$, 
\item the 
ground state wave function for the real Hamiltonian with $\Gamma_d=0$,  
$\psi_{\rm gs}({\rm real})$, 
\end{enumerate}
and
\begin{enumerate}[iii)]
\item the ground state wave function for the elevated-barrier Hamiltonian, 
$\psi_{\rm gs}({\rm eb})$. 
\end{enumerate}
For the case iii), 
we follow the idea of the two-potential method 
\cite{Gurvitz87,Gurvitz88,Gurvitz04}, and modify the 
diagonal energies of the Hamiltonian in the following way: 
we first identify the quadrupole moment at the barrier top, 
$Q_b$, and define 
the maximum diagonal energy of the Hamilotonian at $Q_b$, 
$E_{\rm max}(Q_b)$. We then add 
a constant energy, $E_{\rm add}(Q) = E_{\rm max}(Q_b) - E_{\rm max}(Q)$, 
to the diagonal energies for the configurations at $Q>Q_b$ 
so that the maximum energy becomes a constant 
outside the barrier. 
This is graphically illustrated in Fig. \ref{fig6}
for the case with 3 pairs in 6 orbitals 
(we have found that the results do not 
significantly change even if the energy of the doorway state is set to 
be the same as the lowest diagonal energy at $Q_b$).

\begin{figure} [tb]
\includegraphics[width=1.0\columnwidth,clip]{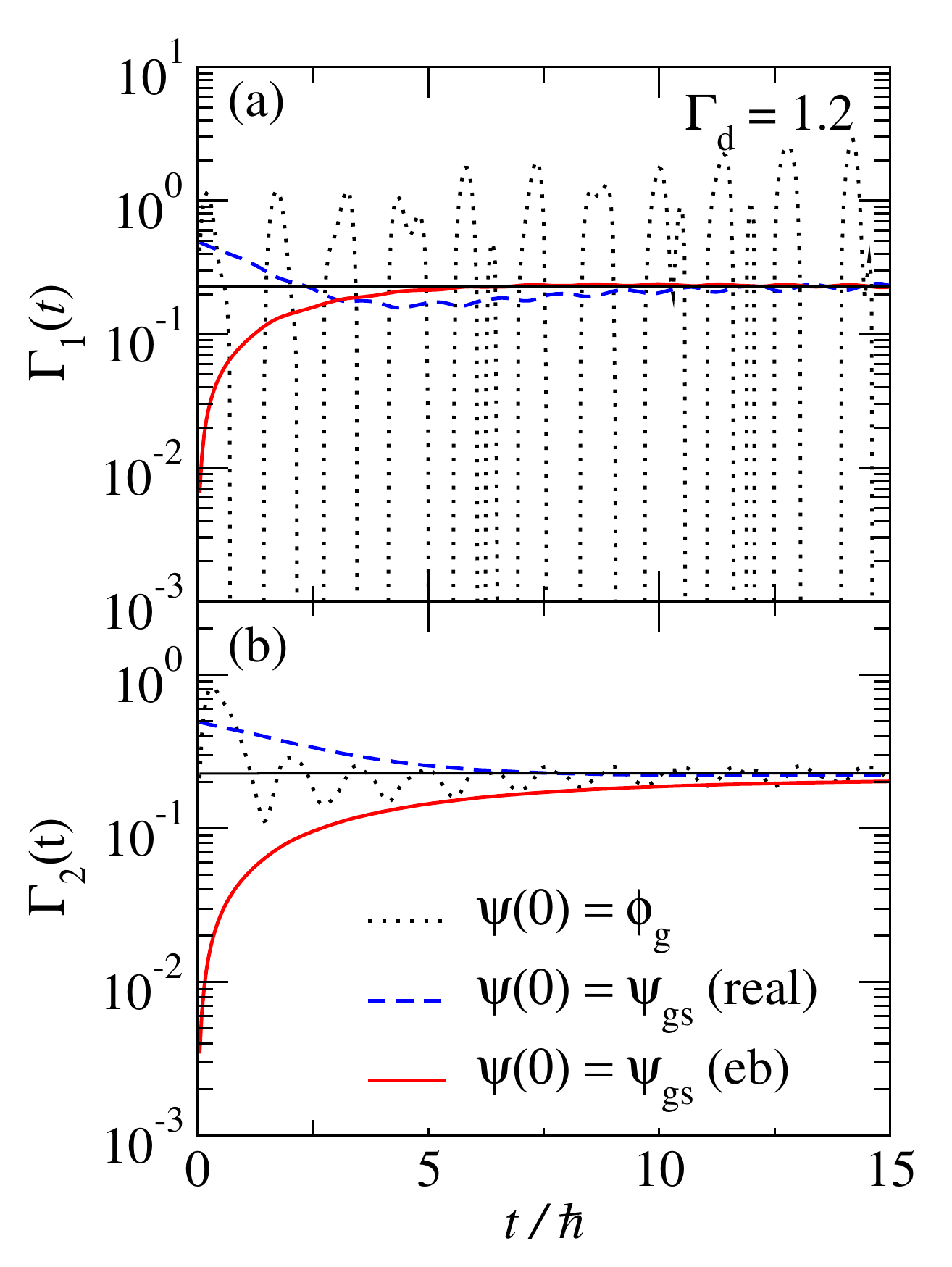}
\caption{
\label{fig7}
The decay widths 
obtained with the time-dependent approach with 
three different initial wave functions. 
These are for the system with 3 pairs in 6 orbitals, with the width of 
the fission doorway configuration of $\Gamma_d=1.2\veps_0$. 
All the energies are measured in 
units of $\veps_0$. 
The upper and the lower panels show the decay widths estimated by 
Eqs. (\ref{gamma1}) and (\ref{gamma2}), respectively. 
The dotted lines use the unperturbed ground state wave function for the 
initial wave function, while the dashed lines are with the ground state 
wave function of the real Hamiltonian with $\Gamma_d=0$. The solid 
lines use the ground state wave function of the modified Hamiltonian 
shown in Fig. \ref{fig6}. The decay width from the time-independent approach 
is denoted by the thin solid lines. 
}
\end{figure}

Figure 7 shows the decay widths obtained by the time-dependent approach 
with the three different initial wave functions. The offset and 
the width for the 
doorway state are set to be $\Delta=0.1\epsilon_0$ and 
$\Gamma_d=1.2\epsilon_0$, respectively. The upper and the 
lower panels show the decay width estimated with 
Eqs. (\ref{gamma1}) and (\ref{gamma2}), respectively. The dotted, the dashed, and 
the solid lines show the results with the choice i), ii), and iii) for the 
initial wave function, respectively. 
One can see that the choice i) does not provide a good result, as the 
decay width is highly oscillating, especially for $\Gamma_1(t)$. 
At large $t$, 
both of the 
choices ii) and iii) lead to the same value of the decay widths as that 
with the time-independent approach, which is denoted by the 
thin solid line. 
This is a natural consequence of the fact that 
only the component with the smallest imaginary energy survives in the 
time-evolution, Eq. (\ref{time-dep}), as $t\to\infty$. 
It is worth noticing that 
the convergence is the fastest for 
$\Gamma_1(t)$ 
with the choice iii), that is, the ground state wave function for a modified 
Hamiltonian. On the other hand, 
the choice ii) would be good when the width for the doorway 
configuration, $\Gamma_d$, is very small, since in this 
case $\psi({\rm real})$ should be almost the same as 
the ground state wave function 
of the non-Hermitian Hamiltonian.

\end{document}